# Towards a Frontier of Spatial Scientometric Studies

SONG GAO

The STKO Lab, Department of Geography, University of California, Santa Barbara, USA
Corresponding email: sgao@geog.ucsb.edu, website: www.geog.ucsb.edu/~sgao/

_______________________________________________________________

The research field of spatial scientometrics is dedicated to measuring and analyzing science with spatial components (e.g., location, place, mapping). It involves the study of spatial patterns, impacts and trends of scientific activities (e.g., co-publication, citation, academic mobility) based on bibliographic databases. Because of the dynamic nature of this field, researchers from interdisciplinary/multidisciplinary domains constantly contribute qualitative, quantitative and computational approaches and technologies into scientometric analysis. This article aims to giving a brief overview about this field by analyzing the publications in (spatial) scientometrics collected from the Scopus database and introduces recent frontier researches which integrate advanced spatial analysis and geovisualization with Semantic Web technologies.

_______________________________________________________________

## 1. INTRODUCTION

The research field of scientometrics (or bibliometrics) is concerned with measuring and analyzing science, to quantify a publication, a journal, or a discipline's structure, impact, change, and interrelations. The spatial dimension (e.g., location, place, distance) of science has been added into account since research activities usually start from a certain region or several places in the world and then spread to other places thus displaying spatiotemporal patterns. The analysis of spatial aspects of the science system is composed of spatial scientometrics [Frenken et al. 2009], which includes the studies of geospatial patterns on scientific activities, domain interactions, co-publications, citations, academic mobility and so forth. The earliest spatial scientometric literature dates back to 1970s. For instance, Frame et al. [1977] analyzed the distribution of world science productivity by region and country. Later on, the availability of more detailed address and coordinate information offers the possibility to investigate the role of physical distance in collaborative knowledge production [Hoekman et al. 2009]. And the "space" could also be "cyberspace" in addition to "physical space". The book *Atlas of Science: Visualizing What We Know* collected amount of visual maps for navigating the dynamic structure of science and technology [Börner 2010].

This article aims to giving a brief overview about the frontier research of this field and introduces several recent spatial scientometric studies which integrate advanced spatial analysis and geovisualization with Semantic Web technologies.

## 2. LITERATURE ANALYSIS

### 2.1 Scientometrics

2. Gao S.

Despite known incompleteness and limitations of bibliographic databases and bibliometric measurements, it can still outline some patterns and trends in related literature for a specified field. We investigated the characteristics of the research field of *Scientometrics* by analyzing the Scopus database, which is one of the largest bibliographic databases containing metadata (e.g., titles, authors, affiliations, abstracts, and citations) of peer-reviewed academic publications all over the world.

The search terms include "scientometrics" or "bibliometrics" which should be mentioned at the article's title, keyword or abstract. When accessing to the database on May 23, 2014, a total number of 6749 articles (including all kinds of document types: journal article, review, editorial, letter, conference paper, book chapter, etc.) were matched with the searching terms. The first publication indexed in this database dates back to 1974 and the number of publications increased gently before 1999 but more sharply after that year (see Figure 1). The scientometric studies help to review and evaluate publication productivity, collaboration, citation impact and topic trends for various disciplines. By analyzing the abovementioned searching results matched with research domains in Scopus, we found that more than 28 domains/subjects have contributed to scientometrics/bibliometrics and the top three contributing domains: *Medicine, Social Science* and *Computer Science* have shared about 60% of the total publications in scientometric studies (see Figure 2). For the country distribution, US ranked the first and followed by UK, Spain and China; the top 20 countries contributed to scientometric research can be found in Figure 3. Those papers have been published on 159 journals; the top 10 publishing journals and the total number of indexed papers were shown in Table I. While *Scientometrics, Journal of the American Society for Information Science and Technology, Malaysian Journal of Library and Information Science, Journal of Informetrics* are domain-specific leading journals for scientometric publications, *Nature, Science and PlOS ONE* also contain many related articles in this subject and might be good source for broader impact across all scientific subjects and domains.

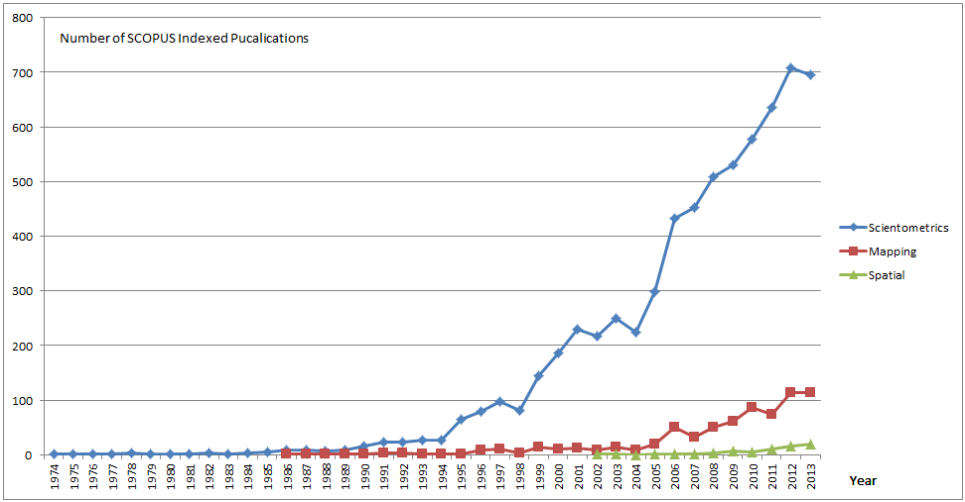

Fig. 1. The temporal patterns of Scopus indexed publications relevant to scientometrics per year.

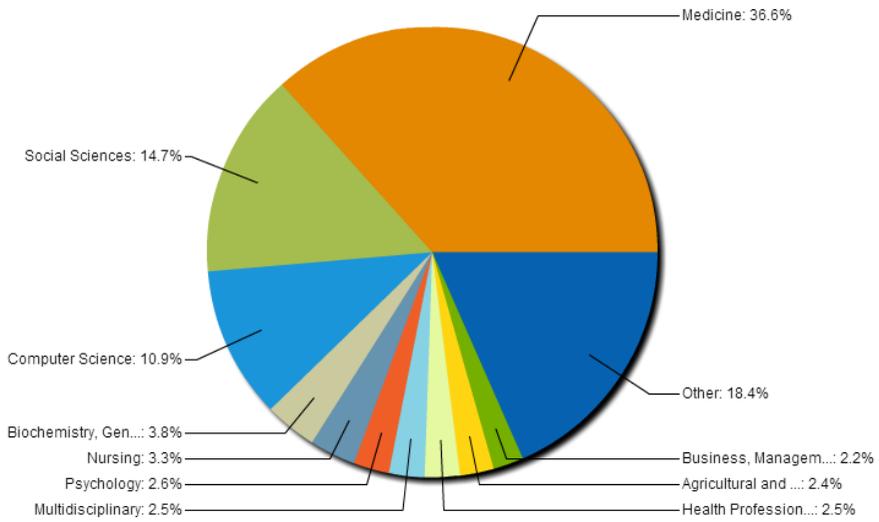

Fig. 2. The share distribution of disciplines contributed to scientometric publications.

4. Gao S.

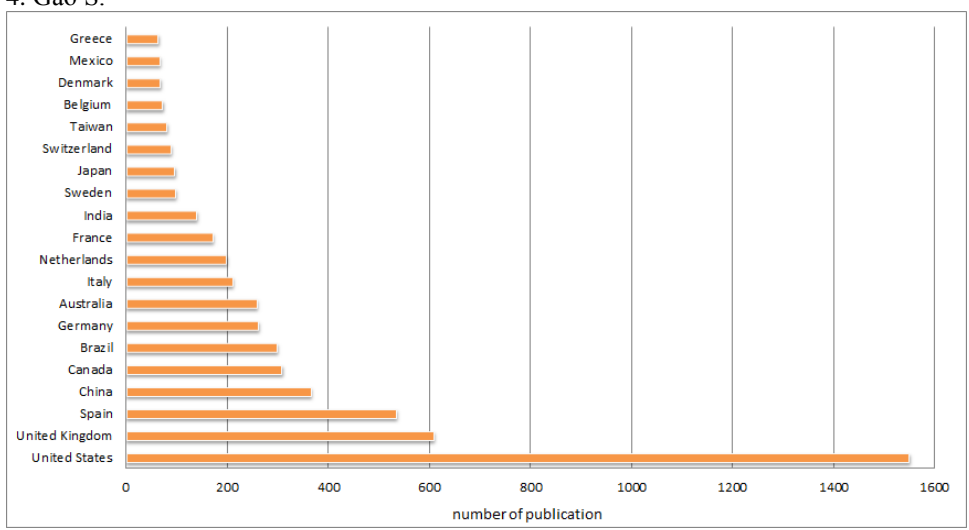

Fig. 3. The distribution of scientometric publications by country

Table I. Top 10 publishing journals and the number of indexed papers related to scientometrics in Scopus

| Publishing Source | Number |
|---|---|
| Scientometrics | 239 |
| Nature | 168 |
| Journal of the American Society for Information Science and Technology | 65 |
| PlOS ONE | 64 |
| Malaysian Journal of Library and Information Science | 58 |
| Journal of Informetrics | 58 |
| Science | 54 |
| Journal of the Medical Library Association | 53 |
| Journal of Clinical Epidemiology | 34 |
| Information Processing and Management | 33 |

## 2.2 Spatial Scientometrics

We refined the previous search results (scientometrics/bibliometrics) with new keywords "mapping" or "spatial" that are popular terms existing in spatial scientometric research. Surprisingly, it only returned 710 articles with "mapping" and 63 with "spatial" from the Scopus database, which approximate 10% of the total publications in scientometric and bibliometric studies. The temporal trends can be compared in Figure 1 and there exists steadily increasing studies in this field. In terms of contributing disciplines to spatial scientometrics (see Figure 4), the top three disciplines are still the same as scientometrics/bibliometrics (see Figure 2), but the rank has changed into *Computer Science, Social Science* and *Medicine*; *Mathematics* rises up to the 4th.

Note that many computer scientists and geographers from a lot of countries have contributed new measurements, analytics and technologies to (geo)spatial scientometric studies with their domain-knowledge and focused on different topics [Frenken et al. 2009]. Figure 5 shows the geospatial distribution of publications related to spatial/mapping scientometrics by country. US, China, UK, Spain, Germany and the Netherland rank at the top countries.

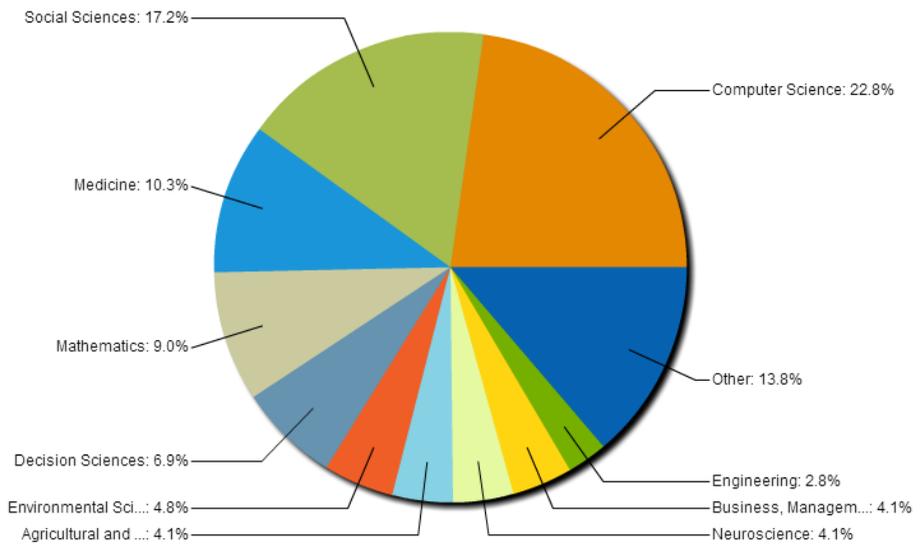

Fig. 4. The share of disciplines contributed to spatial scientometric publications.

6. Gao S.

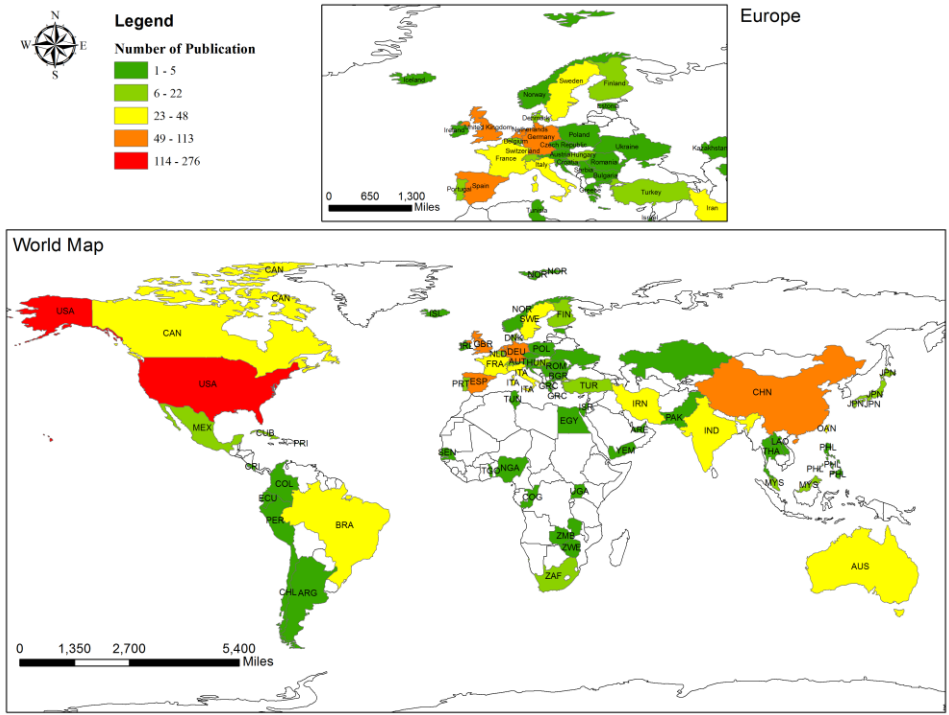

Fig. 5. The geospatial distribution of publications related to spatial/mapping scientometrics by country.

## 3. FRONTIER RESERACH

The above literature analysis demonstrates an overview about scientometric and spatial scientometric studies and publications. In this section, we will introduce several recent scientometric researches which integrate advanced spatial analysis or geovisualization with Semantic Web technologies. An increasing use of Linked Data[1] and SPARQL endpoints has enabled the integration of multiple web sources for publishing, sharing, and interlinking pieces of data, information, and knowledge on the Semantic Web. Thus, complex queries and knowledge discovery can be derived from linked RDF triples. Geospatial information is a crucial part of many central hubs on the Linked Data Web [Janowicz et al. 2012], and usually contained in bibliographic data portals. The geospatial semantics community has started contributing ontologies, analytics and case studies to the spatial scientometric research.

In the field of Geographic Information Science (GIScience), Keßler et al. [2012] introduced a Linked Data portal: *spatial@linkedscience*[2] for exploring the spatio-temporal aspects of bibliography metadata from prominent GIS-domain conference series, including GIScience, COSIT, ACM SIGSPATIAL GIS, and AGILE (see Figure 6). The triple set of <author, affiliation location, time> can help the exploration of individual researcher's trajectory associated with his/her publication in different years.

---

[1] http://linkeddata.org/
[2] http://spatial.linkedscience.org

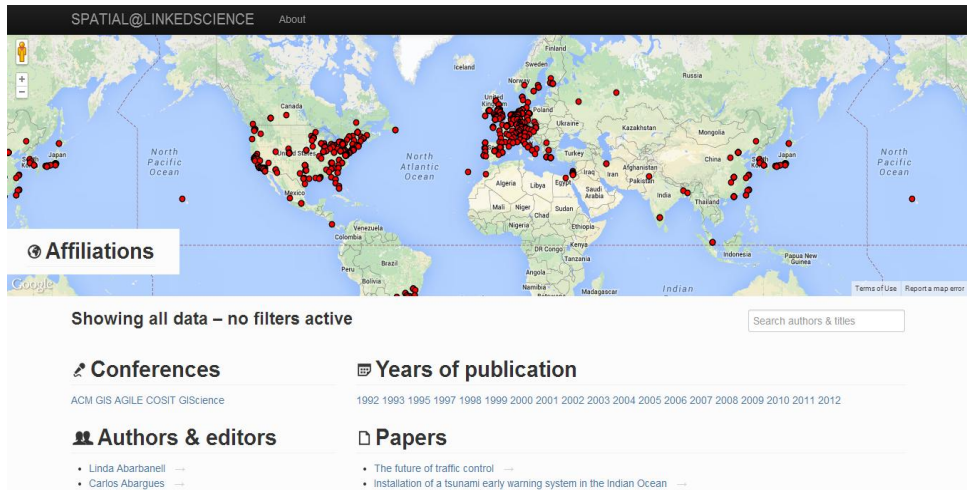

Fig. 6. A screenshot of the Web user interface of spatial@linkedscience.

Another novel research was conducted by Hu et al. [2012]. They designed and developed a Linked-Data-driven and semantically-enabled portal (SWJPortal) [3] for scientometric analysis on *the Semantic Web* journal by IOS Press (Figure 7). This journal follows an open and transparent principle during the whole submitting and review process. It contains almost complete metadata about the publication, timestamp, authors, affiliations, abstracts, reviewers and comments which are available online. Hu et al. [2012] created external links to DBpedia and the Semantic Web Dog Food Corpus to enrich local datasets. Then these data were exposed using a SPARQL endpoint, an extended bibliographic ontology, and a modular Linked Data portal that provides interactive (spatial-)scientometric analytics. The analytic modules include cartogram for visualizing geographic distributions, citation map, co-authorship network and latent dirichlet allocation (LDA) analysis for topic modeling etc., which enables novel (spatial-)scientometric applications, offers insights on scientific networks and discovers new trends. Such a portal can be customized for other journals as well. Recently, this research group further developed another Linked-Data-driven Web portal *DEKDIV*[4] for Learning

---

[3] http://semantic-web-journal.com/SWJPortal/

[4] http://stko-exp.geog.ucsb.edu/lak/

8. Gao S.

Analytics and Knowledge (LAK) and Education Data Mining (EDM) conference series (Figure 8). It was designed to facilitate the exploration, enrichment, visualization, and analysis of the conference bibliography data [Hu et al. 2014]. In order to support geovisualization and reveal geospatial patterns, they geolocated the affiliation of the first author in each publication, reference and citation; These geospatial data are the foundation of spatial scientometric modules such as *Citation Map*, *Reference Map*, *Collaborative Institutes distribution*, and *Mapping Conference Participants*.

The loosely coupled, modular-based Linked-Data-driven infrastructure which has combined multidisciplinary technologies including Semantic Web reasoning, geocoding, D3 visualization library, GeoJSON, and a variety of data mining and machine learning techniques, can be customized and easily migrated to other scientometric projects.

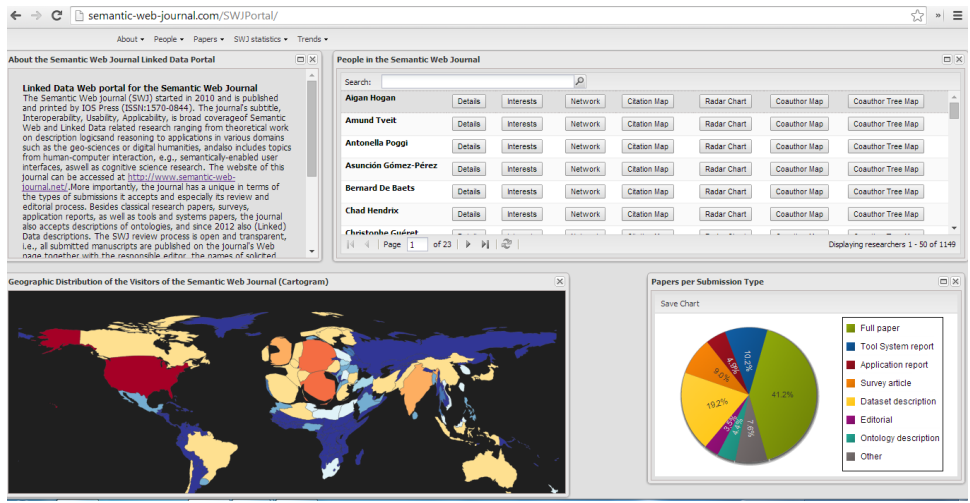

Fig. 7. A screenshot of the Web user interface of the SWJPortal.

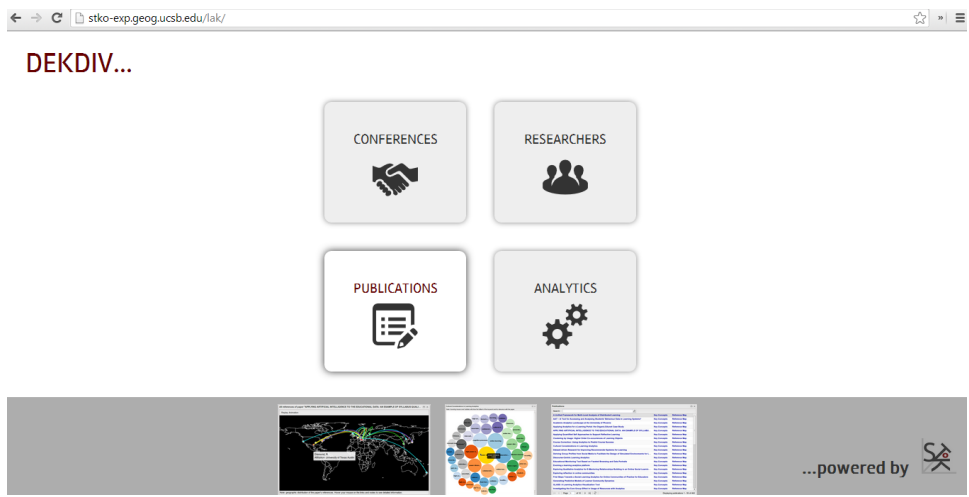

Fig. 8. A screenshot of the Web user interface of the DEKDIV.

The last frontier spatial scientometric research introduced here focused more on spatial-temporal patterns of citations. The coordinate (latitude/longtitude) of an author's affiliation usually was not provided by any bibliography database. By adding coordinate information based on Google Maps Geocoding API and publication metadata from Microsoft Academic Search, Gao et al. [2013] presented a spatiotemporal framework to explore the citation impact of scientific publications as well as individual researchers, with the support of quantitative methods in GIS and spatial statistics. The proposed geospatial s-index for evaluating an individual scientist's geospatial impact may complement traditional non-spatial measures such as h-index and g-index. An interactive Web application *CitationMap*[5] has also been developed for visualizing and understanding how citations spread through space and time.

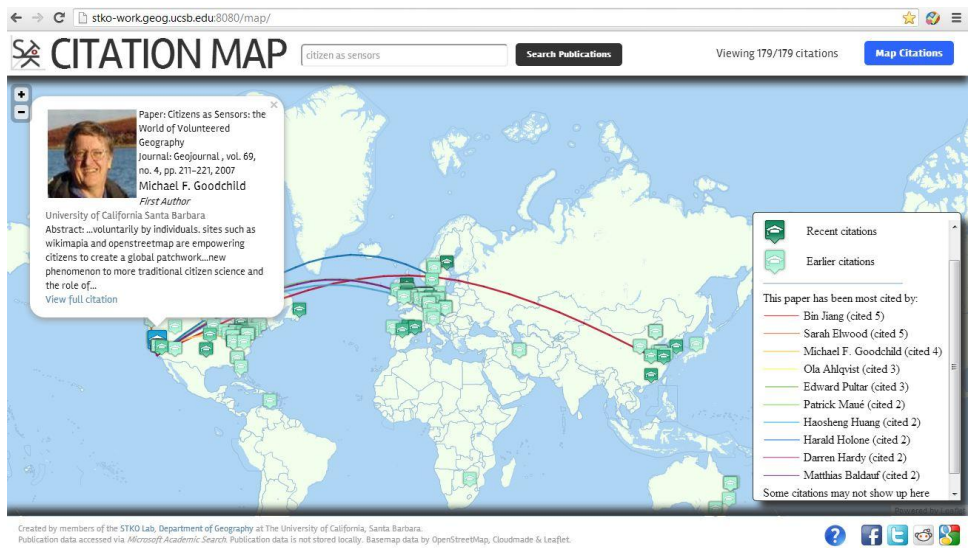

Fig. 9. A screenshot of the Web user interface of the CitationMap

## 4. CONCLUSION

The spatial scientometrics is still an infant multidisciplinary/interdisciplinary field with the support of spatial analysis and statistic methodologies. New spatial measurements to evaluate the excellence in science are emerging. In addition, the increasing use of Semantic Web and mapping technologies yields new insights on spatial scientometric studies. The power of Linked Data Web lies in the capability to enrich and interlink fruitful (un-, semi-)structured data from the Web, which goes beyond the basic bibliography metadata provided by the publishers. Thus, it offers new opportunities for (spatial-)scientometrics to evaluate scientific publications, to investigate the network of authors, reviewers, and editors, or even to predict future research topic trends across domains. The data quality assurance, synthesis and integration issues need more effects in future research.

---

[5] http://stko-work.geog.ucsb.edu:8080/map/

## ACKNOWLEDGMENTS

Many thanks to my colleagues in the STKO lab at UC Santa Barbara for discussions and collaborations in scientometric studies.

_______________________________________________________________________________

Author Bio:

Song Gao is a doctoral student majoring in GIScience and a research assistant in the STKO lab at UC Santa Barbara. His research interests include Big Geo-Data Analytics, Place-based GIS and Advanced Spatiotemporal Analysis. He has served as a peer reviewer for several prominent journals including *Proceedings of the National Academy of Sciences, Transactions in GIS, Computers, Environment and Urban Systems,* and *International Journal of Geographical Information Science.*